\documentclass[12pt]{article}
\usepackage{epsfig, epsf, graphicx, subfigure, color}
\usepackage{pstricks, pst-node, psfrag}
\usepackage{amssymb,amsmath}
\usepackage{verbatim,enumerate}
\usepackage{rotating, lscape}
\usepackage{setspace}
\usepackage{bbm}
\usepackage{natbib}
\usepackage{amsthm}
\usepackage{url}
\usepackage{appendix}

\theoremstyle{definition}
	
\newtheorem{Definition}{Definition}
\newtheorem{Theorem}{Theorem}

\newcommand{\bi}{\begin{itemize}}
	\newcommand{\ei}{\end{itemize}}

\begin{document}
	\thispagestyle{empty} \baselineskip=28pt \vskip 5mm
	\begin{center} {\Large{\bf Total Variation Depth for Functional Data}}
	\end{center}
	
	\baselineskip=12pt \vskip 5mm
	
	\begin{center}\large
		Huang Huang\footnote[1]{
			\baselineskip=10pt CEMSE Division, King Abdullah University of Science and Technology, Thuwal 23955-6900, Saudi Arabia.
			E-mail: huang.huang@kaust.edu.sa; ying.sun@kaust.edu.sa
		} and Ying Sun$^1$\end{center}
	
	\baselineskip=16pt \vskip 1mm \centerline{\today} \vskip 8mm
	
	\begin{center}
		{\large{\bf Abstract}}
	\end{center}
	\baselineskip=17pt
				
			There has been extensive work on data depth-based methods for robust multivariate data analysis. Recent developments have moved to infinite-dimensional objects such as functional data. In this work, we propose a new notion of depth, the total variation depth, for functional data. As a measure of depth, its properties are studied theoretically, and the associated outlier detection performance is investigated through simulations. Compared to magnitude outliers, shape outliers are often masked among the rest of samples and harder to identify. We show that the proposed total variation depth has many desirable features and is well suited for outlier detection. In particular, we propose to decompose the total variation depth  into two components that are associated with shape and magnitude outlyingness, respectively. This decomposition allows us to develop an effective procedure for outlier detection and useful visualization tools, while naturally accounting for the correlation in functional data. Finally, the proposed methodology is demonstrated using real datasets of curves, images, and video frames.	
	\begin{doublespace}
		
		\par\vfill\noindent
		{\bf Some key words}: data depth, functional data, total variation, outlier detection, shape outliers
		\par\medskip\noindent
		{\bf Short title}: Total Variation Depth
	\end{doublespace}
	
	\clearpage\pagebreak\newpage \pagenumbering{arabic}
	\baselineskip=26.5pt
	

\section{Introduction}\label{sec:intro}
Functional data, realizations of a one-dimensional stochastic process, in the form of functions, are observed and collected with increasing frequency across research fields, including meteorology, neuroscience, environmental science and engineering. Functional data analysis (FDA) considers the continuity of functions and various parametric and nonparametric methods can be found in \cite{Ferraty2006} and \cite{Ramsay2009}. 
In recent years, extensive developments have extended typical techniques of FDA to the analysis of more complicated functional objectives. Besides model-based methods, exploratory data analysis (EDA) has been extended to functional data as well. 

\cite{Sun2011} proposed the functional boxplot as an informative visualization tool for functional data, and \cite{Genton2014} extended it to image data. Similar to the classical boxplot, if we simply extend the univariate ranking to the functional setting, the features of functional data cannot be captured. Data depth is a widely used concept in multivariate and functional data ranking. The general requirement for a data depth notion is the so-called ``center-outwards'' ordering, which means the natural center of the functional data should have the largest depth value and the depth decreases as the data approach outwards. \cite{Liu1999} reviewed many popular depth notions for multivariate data. Some examples in the nonparametric framework include half-space depth~\citep{Tukey1975}, simplicial depth~\citep{Liu1990}, projection depth~\citep{Zuo2000}, and spatial depth~\citep{Serfling2002}. For univariate functional data depth,  (modified) band depth ~\citep{Lopez-Pintado2009}, integrated data depth~\citep{Fraiman2001}, half-region depth~\citep{Lopez-Pintado2011}, and extremal depth~\citep{Narisetty2015} have been proposed depending on the desired emphases.
Nowadays, the study of multivariate functional data also arouses extensive interests due to its practical applications. \cite{Berrendero2011} studied the daily temperature functions on different surfaces, and \cite{Sangalli2009} and \cite{Pigoli2012} analyzed multivariate functional medical data. To provide a valid depth of multivariate functional data, \cite{Ieva2013} derived depth measures for multivariate
functional data from averaging univariate functional data depth, and
\cite{Claeskens2014} gave a generalization to multivariate functional data depths by averaging a multivariate depth function over the time points with a weight function. Particularly, \cite{Lopez-Pintado2014} proposed and studied the simplicial band depth for multivariate functional data, which is an extension of the univariate functional band depth.

Data depth-based methods provide many attractive tools in solving problems related to classification, clustering, and outlier detection. For example, \cite{Jornsten2004}, \cite{Ghosh2005} and \cite{Dutta2012} used various depth notions to classify or cluster multivariate data. \cite{lopez2006} and \cite{Cuevas2007} extended the classification problem to functional data. 
Outlier detection is another challenging problem, especially for functional data, because there is no clear definition of functional outliers.
Roughly speaking, functional outliers can be categorized into two types: magnitude and shape outliers. Magnitude outliers have very deviated values in some dimensions, while shape outliers have a different shape compared to the vast majority; however, no data value may deviate too much from the center.
\cite{Dang2010} introduced nonparametric multivariate outlier identifiers based on various multivariate depths.
For functional data, \cite{Febrero2008} used a cutoff, which is determined by a bootstrap, for the functional data depth to detect outliers.
\cite{Hyndman2010} proposed using the first two robust principal components to construct a bagplot~\citep{Rousseeuw1999}, or a highest density region plot~\citep{Hyndman1996} to detect outliers. However, the first two principle components often do not adequately describe the variabilities in functional data. In the functional boxplot proposed by \cite{Sun2011}, outliers were detected by the $1.5$ times the $50\%$ central region rule, which means that any departure of an observation from the fence, that is, the inflation of the $50\%$ central region by $1.5$ times, makes it an outlier. The $50\%$ central region is the envelope of the first $50\%$ of curves that have the largest depth values, and the factor $1.5$ can be adjusted according to the distribution~\citep{Sun2012}. Good performance has been shown for various types of outliers, especially for magnitude outliers. However, for shape outliers, two issues arise: outliers may appear among the first $50\%$ of curves with the largest depth values and outliers that are masked in the fence can never be detected, even though their depth values are small. Recently, \cite{Arribas-Gil2014} proposed the outliergram, where the relationship between modified band depth~\citep{Lopez-Pintado2009} and modified epigraph index~\citep{Lopez-Pintado2011} is studied to identify shape outliers. \cite{Hubert2015} proposed functional bagdistance and skewness-adjusted projection depth to detect various outliers for multivariate functional data.

In this paper, we propose a new notion of data depth, the total variation depth, and develop an effective procedure to detect both magnitude and shape outliers using attractive features of the total variation depth. We were motivated by the drawback of the modified band depth. Typically, functional data depth is constructed via averaging the pointwise depth. The modified band depth can be viewed as such an average for temporal curves, however, it does not take the time order or temporal correlations into account. In contrast, our proposed total variation depth allows for a meaningful decomposition that considers the correlation of adjacent dimensions and can be used for outlier detection. 
In our proposed outlier detection procedure, we show that combined with the functional boxplot, we are able to detect both magnitude and shape outliers. We also develop informative visualization tools for easy interpretation of the outlyingness. The outlier detection performance has been examined through simulation studies. The visualization tool is described and illustrated through three kinds of real-data applications: curves, images and video frames.

\section{Methodology}
\subsection{Total variation depth}
\label{subsec:TVD}
Let $X$ be a real-valued stochastic process on $\mathcal{T}$ with distribution $F_X$, where $\mathcal{T}$ is an interval in $\mathbb{R}$. We propose the total variation depth for a given function $f$ w.r.t.~$F_X$. We use $f$ to denote a function, and $f(t)$ to denote the functional value at a given $t$.
First, we define the pointwise total variation depth for a given $t$. Let $R_{f}(t)=\mathbbm{1}\{ X(t)\leq f(t)\},$ where $\mathbbm{1}$ is the indicator function. 
It is easy to see that $p_f(t)=\mathbb{E}\{R_f(t)\}=\mathbb{P}\{X(t)\leq f(t)\}$ is associated with the relative position of $f(t)$ w.r.t.~$X(t)$. 
For instance, if $f^\star(t)$ is the true median, then $p_{f^\star}(t)=1/2$. 
The pointwise variation depth is defined as follows:
\begin{Definition}
	\label{def:pointTVD}
	For a given function $f(t)$ at each fixed $t$, the pointwise total variation depth of $f(t)$ is defined as $D_f(t)$,
	$$D_f(t)=\hbox{Var}\{R_{f}(t)\}=p_f(t)\{1-p_f(t)\}.$$
\end{Definition}
For a fixed $t$, $D_f(t)$ is maximized at the center when $p_f(t)=1/2$, or simply the univariate median.
Next, we define the functional total variation depth (TVD) for the given function $f(t)$ on $\mathcal{T}$. 
\begin{Definition}
	\label{def:TVD}
	Let $D_f(t)$ be the pointwise total variation depth from Definition~\ref{def:pointTVD}. The TVD for function $f(t)$ on $\mathcal{T}$ is given by
	\[\hbox{TVD}(f)=\int_\mathcal{T} w(t)D_{f}(t)dt,\]
	where $w(t)$ is a weight function defined on $\mathcal{T}$.
\end{Definition}
There are many ways to choose the weight function, and we now provide two examples for the choice of $w(t)$. If we let $w(t)$ be a constant, $w_1(t)\equiv1/\mid \mathcal{T} \mid$, then it can be shown that the TVD is equivalent to the modified band depth~\citep{Lopez-Pintado2009}. Another example is to choose the weight according to the variability of $X(t)$ at different time points. For instance, similar to \cite{Claeskens2014}, we let $w_2(t)=\hbox{sd}\{X(t)\}/\int_{\mathcal{T}}\hbox{sd}\{X(s)\}ds$, where $\hbox{sd}$ stands for the standard deviation. Intuitively, assigning more weight to time points where sample curves are more spread out will lead to a better separation in the depth values of these sample curves. We choose $w_2(t)$ for the study hereinafter in this paper.

\subsection{Properties of the total variation depth}
\label{subsec:property}
\cite{Zuo2000} studied the key properties of a valid multivariate data depth, and \cite{Claeskens2014} extended them to a functional setting, including affine invariance, maximality at the center, monotonicity relative to the deepest point, and vanishing at infinity. 
By denoting the total variation depth of $f$ w.r.t. $F_X$ by $\hbox{TVD}(f,F_X)$, we first show that it enjoys these desirable properties as a depth notion in Theorem~\ref{theorem:property}. Then, in Theorem~\ref{theorem:decomposition}, we show that the pointwise total variation depth can be decomposed into the pointwise magnitude variation and pointwise shape variation, which has important practical implications for outlier detection problems. We omit the proof of Theorem~\ref{theorem:property}, since it is straightforward. The decomposition in Theorem~\ref{theorem:decomposition} follows immediately from the law of the total variance.
\begin{Theorem}[\nopunct {\bf Properties of TVD}]
	\label{theorem:property}
	 The total variation depth $\hbox{TVD}(f,F_X)$ defined in Definition~\ref{def:TVD} satisfies the following properties.
	\begin{itemize}
		\item Affine invariance. $\hbox{TVD}(f,F_X)=\hbox{TVD}(af+g,F_{aX+g})$ for any $a\in\mathbb{R}\backslash\{0\}$ and any function $f$ and $g$ on $\mathcal{T}$.
		\item Maximality at the center. If at each time point $t\in \mathcal{T}$, the distribution $F_X(t)$ has a uniquely defined center $f^\star(t)$, then $\hbox{TVD}(f^\star,F_X)=\sup_{f\in C(\mathcal{T})}\hbox{TVD}(f,F_X)$, where $C(\mathcal{T})$ denotes the set of all continuous functions on $\mathcal{T}$.
		\item Monotonicity relative to the deepest point. If $f^\star$ is the function such that $\hbox{TVD}(f^\star,F_X)=\sup_{f\in C(\mathcal{T})}\hbox{TVD}(f,F_X)$, then $\hbox{TVD}(f,F_X)\leq\hbox{TVD}(f^\star+\alpha(f-f^\star),F_X)$, for any $f$ on $\mathcal{T}$ and $\alpha\in[0,1]$.
		\item Vanishing at infinity. $\hbox{TVD}(f,F_X)\to 0$ as $\|f(t)\|\to 0$ for almost all time points $t$ in $\mathcal{T}$. Particularly, if the weight function $w(t)=\hbox{sd}\{X(t)\}/\int_{\mathcal{T}}\hbox{sd}\{X(s)\}ds$ is used, then $\hbox{TVD}(f,F_X)\to 0$ as $\|f\|\to 0$, which is the null at infinity property stated by \cite{Mosler2012}.
	\end{itemize}
\end{Theorem}
\begin{Theorem}[\nopunct {\bf Decomposition of TVD}]
	\label{theorem:decomposition}
	Let $s,t$ be two time points satisfying $s\leq t$. The pointwise total variation depth has the following decomposition: \[D_f(t)=\hbox{Var}\{R_{f}(t)\}=\hbox{Var}[\mathbb{E}\{R_f(t)\mid R_f(s)\}]+\mathbb{E}[\hbox{Var}\{R_f(t)\mid R_f(s)\}].\]
\end{Theorem}

The decomposition implies that the total variance of $R_f(t)$ can be decomposed into $\hbox{Var}[\mathbb{E}\{R_f(t)\mid R_f(s)\}]$, the variability that is explained by $R_f(s)$, and $\mathbb{E}[\hbox{Var}\{R_f(t)\mid R_f(s)\}]$, the variability that is independent of $R_f(s)$. We call $\hbox{Var}[\mathbb{E}\{R_f(t)\mid R_f(s)\}]$ the shape component and $\mathbb{E}[\hbox{Var}\{R_f(t)\mid R_f(s)\}]$ the magnitude component. If the shape component of a sample curve dominates the total variation depth, it indicates that the curve has a similar shape compared to the majority. Now we give the definition of the shape variation using the ratio of the shape component to the total variation depth.
\begin{Definition}
	\label{def:SV}
	For any function $f$ on $\mathcal{T}$, the shape variation (SV) of $f$ w.r.t. the distribution $F_X$ is defined as
	\[\hbox{SV}(f)=\int_\mathcal{T}v\{t;s(t)\}S_f\{t;s(t)\}dt,\]
	where $v\{t;s(t)\}$ is a weight function, and at each time point $t$, $s(t)$ is a chosen time point in $\mathcal{T}$ satisfying $s(t)\leq t$, and $S_f\{t;s(t)\}$ is given by 
	\[
	S_f\{t;s(t)\}=\left\{\begin{array}{lcr}
	\hbox{Var}\big(\mathbb{E}[R_f(t)\mid R_f\{s(t)\}]\big)/D_f(t)&,& D_f(t)\neq0\\
	1&,& D_f(t)=0\\
	\end{array}\right..
	\]
\end{Definition}
Now we discuss the choice of the weight function $v\{t;s(t)\}$. Notice that at a given $t$, $R_{f}(t)=\mathbbm{1}\{ X(t)\leq f(t)\}$ is an indicator function. If $X(t)\leq f(t)$, $R_f(t)=1$ no matter how large the value of $f(t)$ is. To account for the value of $f(t)$, we choose the weight function $v\{t;s(t)\}$ to be the normalized changes in $f(t)$ on $\mathcal{T}$. More precisely, $v\{t;s(t)\}$ is given by $v\{t;s(t)\}=\mid f(t)-f\{s(t)\}\mid/\int_{\mathcal{T}}\mid f(t)-f\{s(t)\}\mid$. Then, more weights are assigned to the time intervals where $f(t)$ has larger changes in magnitude. This choice of $v\{t;s(t)\}$ is useful in practice when sample curves are only different in magnitude within short time intervals.

In Definition~\ref{def:SV}, smaller values of the shape variation are associated with larger shape outlyingness. However, we notice that for outlying pairs $(f\{s(t)\},f(t))$ with a small value of the shape component $\hbox{Var}\big(\mathbb{E}[R_f(t)\mid R_f\{s(t)\}]\big)$, $S_f(t;s(t))$ may not be small enough, if $D_f(t)$ in the denominator is too small. To better reflect the shape outlyingness via the shape variation, we shift $(f\{s(t)\},f(t))$ to the center, such that $(\tilde f\{s(t)\},\tilde f(t))=(f\{s(t)\},f(t))-\triangle_t$, where $\triangle_t=f(t)-\hbox{median}\{X(t)\}$, and then define the modified shape variation using the shifted pairs as follows:
\begin{Definition}
	For any function $f$ on $\mathcal{T}$, the modified shape variation (MSV) of $f$ w.r.t. the distribution $F_X$ is defined as
	\[\hbox{MSV}(f)=\int_\mathcal{T}v\{t;s(t)\}S_{\tilde f}\{t;s(t)\}dt.\]
\end{Definition}
Similar to the shape variation, the modified shape variation characterizes shape outlyingness of $f$ via each pair of function values, but it is a stronger indicator by shifting each pair to the center when calculating $S_{\tilde f}\{t;s(t)\}$. \cite{Arribas-Gil2014} used a similar idea for shape outlier detection, whereas the entire extreme curve at all time points is shifted to the center by the same amplitude.

In real applications, we observe functional data at discretized time points and do not know the true distribution. We replace the cumulative density function $F_X$ by its empirical version and obtain 
the sample total variation depth and the sample modified shape variation. All the estimations use sample proportions to approximate probabilities, and the details are provided in the Appendix. Next, we show the consistency of the sample total variation depth, denoted by $\widehat{\hbox{TVD}}$.

\begin{Theorem}[\nopunct {\bf Consistency of TVD}]
	\label{theorem:consistency}
	 Let $\mathcal{S}=\{X_1(t), \ldots, X_n(t)\}$ be the $n$ samples of the stochastic process $X(t)$ observed at $\{t_1<t_2<\ldots<t_m\}.$ Let $h_m(t)$ be the design density with $H(t)=\int_{-\infty}^{t}h_m(u)du$, such that $t_i=H^{-1}(\frac{i-1}{m-1})$. If the following conditions are satisfied	
	\begin{itemize}
		\item[(C1)] $\mathbb{E}\{X(t)\}<\infty$;
		\item[(C2)] $\widehat{\mathcal{T}}=[H^{-1}(0),H^{-1}(1)]$ is compact;
		\item[(C3)] $h_m(t)$ is differentiable;
		\item[(C4)] $\inf_{t\in\widehat{\mathcal{T}}}h_m(t)>0$;
		\item[(C5)] $\int_{\widehat{\mathcal{T}}}|w(t,F_{\tilde X})-w(t,F_{X})|dt\rightarrow0,$ as $n\rightarrow\infty$, where $w(t,F_X)$ denotes the weight function with a specific distribution $F_X(t)$ at time $t$, and $F_{\tilde X}(t)$ denotes the distribution of $\tilde X$, an interpolating continuous process with details in the Appendix; and
		\item[(C6)] the distribution $F_X(t)$ at time $t$ is absolutely continuous with bounded density;
	\end{itemize}
	then
	\[
	\sup_{f\in C(\widehat{\mathcal{T}})} |\hbox{TVD}(f)-\widehat{\hbox{TVD}}(f)|\to 0,~~~~~~ a.s.,
	\]
	as $m\to\infty$ and $n\to\infty$.
\end{Theorem}
When proving the consistency of the sample multivariate functional depth, conditions C1--C6 follow \cite{Claeskens2014}. They are mild conditions, and they are easy to verify for a given stochastic process. Specifically, C1 is a natural condition for a stochastic process, C2--C4 are the conditions for the design distribution of the observation locations, and C5 is the condition for the weight function. Using similar arguments as in \cite{Claeskens2014}, it is easy to show that the weight functions $w_1(t)$ and $w_2(t)$ propsoed in \S\ref{subsec:TVD} satisfy C5; and C6 characterizes the property of the distribution of $X$ marginally at each location. A brief proof of Theorem~\ref{theorem:consistency} is provided in the Appendix.

\subsection{Outlier detection rule and visualization}
\label{subsec:outlierDetectRule}
To obtain robust inferences, outlier detection is often necessary; however, the procedure is challenging for functional data, because the characterization of functional data in infinite dimensions and appropriate outlier detection rules are needed. There are many proposed ways to detect functional outliers, some of which have been discussed in \S\ref{sec:intro}, and here, we propose a new outlier detection rule by making good use of the decomposition property of the proposed total variation depth. Suppose we observe $n$ sample curves, then the outlier detection procedure is summarized as follows:
\begin{enumerate}
	\item Estimate the total variation depth and modified shape variation for each curve as described in \S\ref{subsec:TVD} and \S\ref{subsec:property}.
	\item Draw a classical boxplot for the $n$ values of the modified shape variation and detect outliers by the $3\times\hbox{IQR}$ empirical rule, where the factor $3$ can be adjusted by users when necessary. Curves with modified shape variation values below the lower fence in the boxplot are identified as  shape outliers.
	\item Remove detected shape outliers and draw a functional boxplot using the total variation depth to detect all the magnitude outliers by $1.5$ times of the $50\%$ (w.r.t. the number of original observation before removing shape outliers) central region rule. The factor $1.5$ can also be adjusted~\citep{Sun2012}.
\end{enumerate}

It is worth noting that using the functional boxplot along with the boxplot of the modified shape variation enables us to detect both magnitude outliers and shape outliers with small oscillations. We call the boxplot of the modified shape variation the shape outlyingness plot because the modified shape variation is a good indicator of shape outlyingness. It is useful especially in detecting shape outliers without a significant magnitude deviation.
\begin{figure}[ht!]
	\begin{center}
		\includegraphics[width=0.95\textwidth]{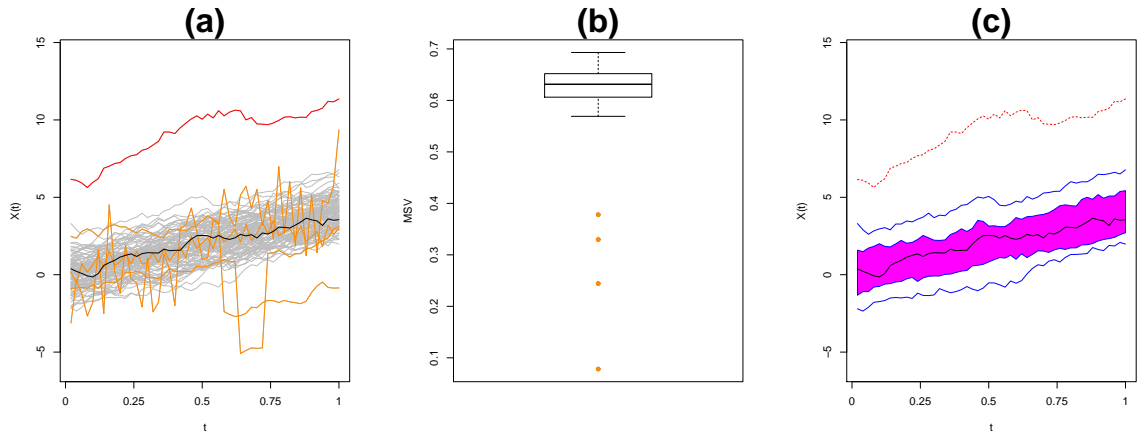}
		\caption{\label{fig:visualization} Visualization tools: (a) Observations with a detected magnitude outlier (red) and shape outliers (orange). (b) Shape outlyingness plot with points in orange corresponding to shape outliers. (c) Functional boxplot using the total variation depth after removing detected shape outliers.}
	\end{center}
\end{figure}

Next, we propose a set of informative visualization tools following one outlier detection procedure.
Fig.~\ref{fig:visualization}~(a) shows three examples of simulated datasets containing different types of outliers. The model details will be introduced in \S\ref{subsec:outlierModel}. The grey curves are the non-outlying observations and the black curve is the functional median with the largest total variation depth value. The highlighted orange curves are the shape outliers detected by the shape outlyingness plot, as shown in Fig.~\ref{fig:visualization}~(b), where the points in orange correspond to shape outliers.
The red curve is a magnitude outlier detected by the functional boxplot constructed in Step 3 of the outlier detection procedure after removing the shape outliers. Finally, the functional boxplot in Fig.~\ref{fig:visualization}~(c) displays the median, the $50\%$ central region, the maximal and minimal envelopes, and the magnitude outlier.

\section{Simulation Study}
\label{sec:simulation}
\subsection{Outlier models}
\label{subsec:outlierModel}
In this section, we choose a similar simulation design to that by ~\cite{Sun2011} and \cite{Narisetty2015}, but we also introduce new outlier models. We study seven models in total, where Model 1 is the base model with no outliers compared to Model 2~--6, and Model 7 concerns another kind of contamination with a different base model. The contamination ratio for outliers in Model 2--7 is $\epsilon=10\%$. For all the models, $\mathcal{T}$ is set to be $\mathcal{T}=\{t:t\in[0,1]\}$.

{\bf Model 1:} $X_i(t)=4t+e_i(t)$, for $i=1,\ldots,n,$ where $e_i(t)$ is a zero-mean Gaussian process with covariance function $c(s,t)=\hbox{exp}\{-|s-t|\}$.

{\bf Model 2:} $X_i(t)=4t+e_i(t)+6c_i\sigma_i$, for $i=1,\ldots,n$ where $c_i\sim$Bernoulli$(\rho)$ and $\sigma_i$ takes values of $1$ and $-1$ with probability $1/2$, respectively. 

{\bf Model 3:} $X_i(t)=4t+e_i(t)+6c_i\sigma_i,$ if $t\geq T_i,$ and $X_i(t)=4t+e_i(t),$ if $t<T_i,$ for $i=1,\ldots,n$, where $T_i\sim$Unif$([0,1])$, and $c_i$ and $\sigma_i$ have the same definition as in Model 2.

{\bf Model 4:} $X_i(t)=4t+e_i(t)+6c_i\sigma_i,$ if $T_i\leq t\leq T_i+l,$ and $X_i(t)=4t+e_i(t)$ otherwise, for $i=1,\ldots,n$, where $l=0.08, T_i\sim$Unif$([0,1-l])$, and $c_i$ and $\sigma_i$ have the same definition as in Model 2.

{\bf Model 5:} $X_i(t)=4t+(1-c_i)e_i(t)+c_i\tilde e_i(t),$ for $i=1,\ldots,n$, where $c_i$ and $e_i(t)$ have the same definition as before, and $\tilde e_i(t)$ is another zero-mean Gaussian process with a different covariance function $\tilde c(s,t)=6\hbox{exp}\{-|s-t|^{0.1}\}$.

{\bf Model 6:} $X_i(t)=4t+e_i(t)+c_i\big(0.5\sin(40\pi t)\big),$ for $i=1,\ldots,n$, where $c_i$ and $e_i(t)$ have the same definition as before.

{\bf Model 7:} $X_i(t)=2\sin(15\pi x+2c_i)+e_i(t),$ and the corresponding base model in Model 7 is $X_i(t)=2\sin(15\pi x+2c_i)+e_i(t)$, for $i=1,\ldots,n$,  where $c_i$ and $e_i(t)$ have the same definition as before.

\begin{figure}[ht!]
	\begin{center}
		\includegraphics[width=0.8\textwidth]{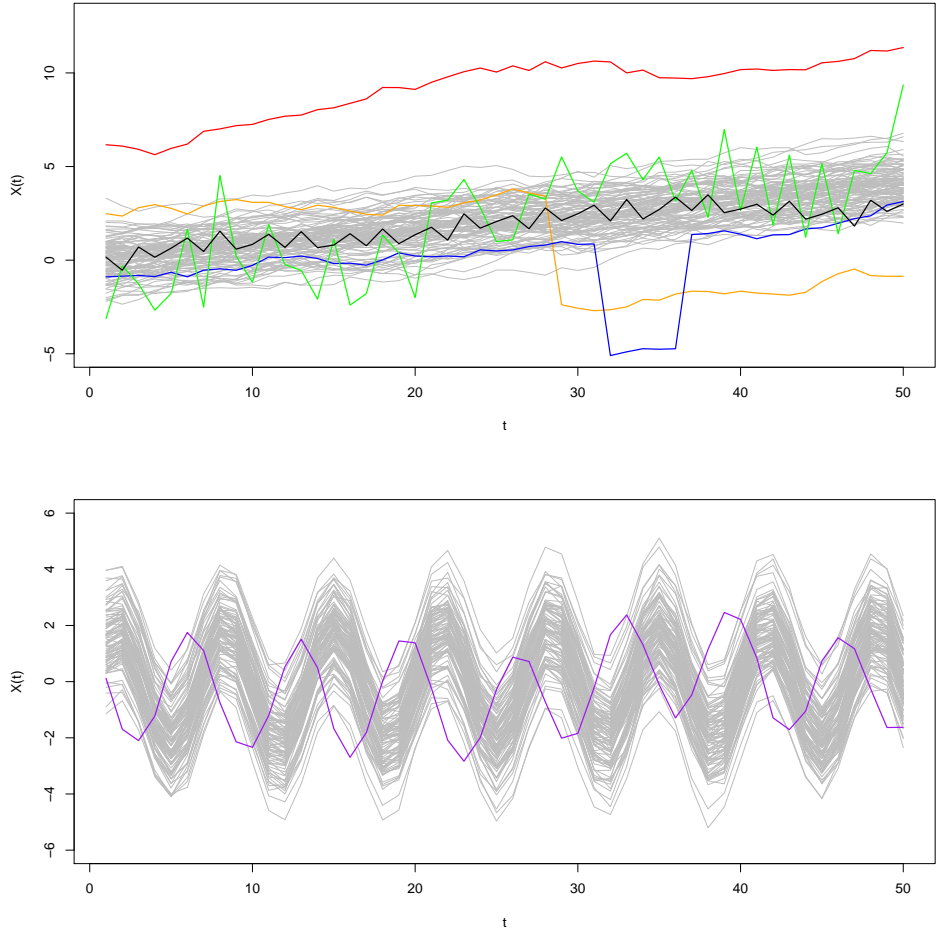}
		\caption{\label{fig:dataset}Top panel: 100 curves (grey) from Model 1 with one of different outliers from Model 2 (red), Model 3 (orange), Model 4 (blue), Model 5 (green) and Model 6 (black), respectively. Bottom panel: 100 curves (grey) contaminated by one outlier from Model 7 (purple).}
	\end{center}
\end{figure}

Then, Model 2 contains symmetric magnitude outliers with a shift; Model 3 makes the magnitude outliers deviate starting from a random time point; Model 4 generates magnitude outliers that have peaks lasting for a short time period; Model 5 has outliers that have a different temporal covariance, so that the outliers have a different shape as well as a larger variance; Model 6 considers shape outliers by adding an oscillating function, where the oscillation is frequent in time but close to the majority in magnitude; and Model 7 introduces shape outliers with a phase shift. The illustration of generated outliers from Model 2--7 is shown in Fig.~\ref{fig:dataset}. For the following simulation studies, we generate $n=100$ curves taking values on an equally spaced grid of $[0,1]$ with $p=50$ time points for each model.

\subsection{Central region}
\begin{figure}[p!]
	\begin{center}
		\includegraphics[width=\textwidth]{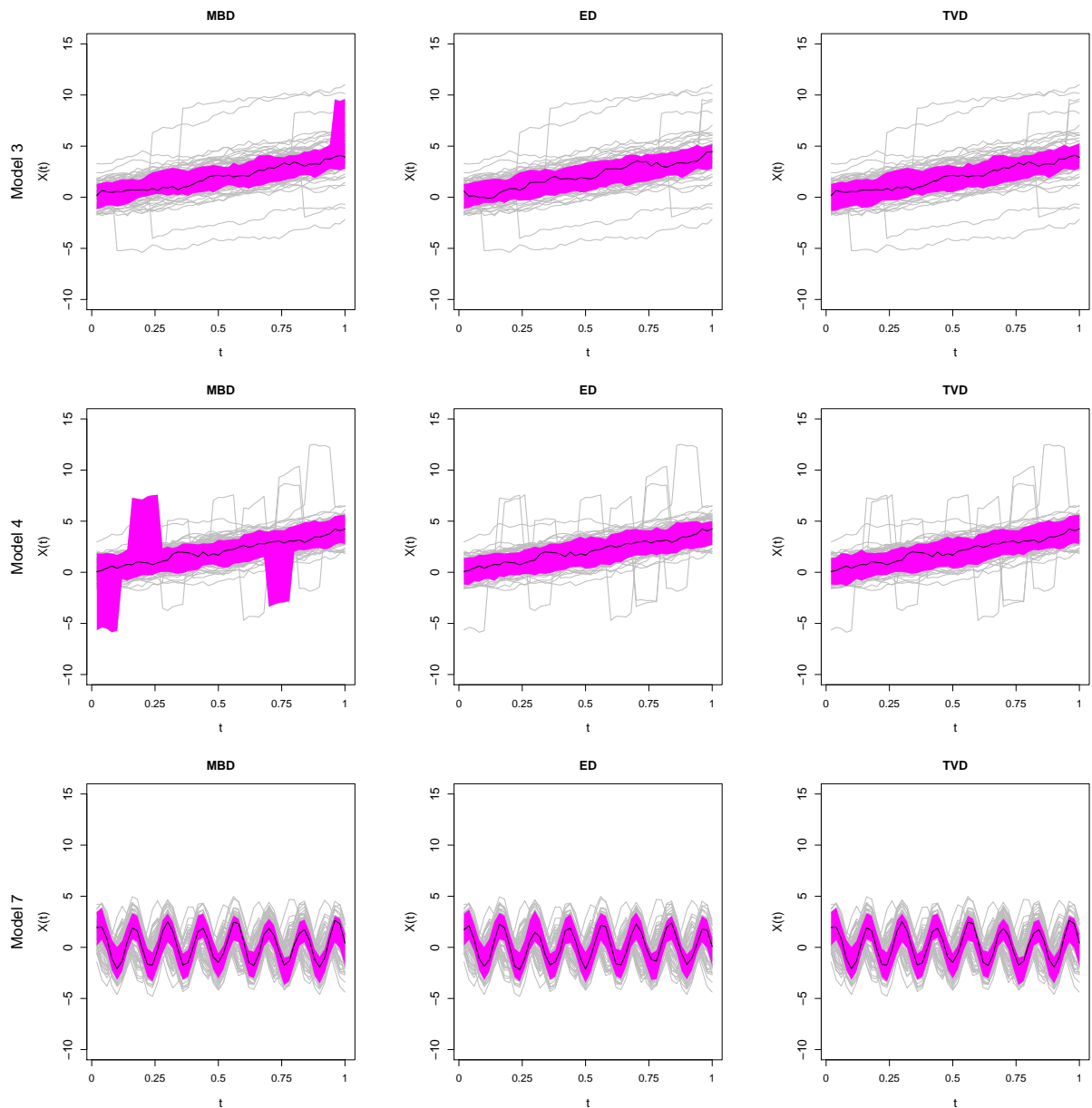}
		\caption{\label{fig:centralRegion}The central regions constructed by MBD (left), ED (middle), and TVD (right) for Model 3 (top), Model 4 (middle) and Model 7 (bottom). The central region is displayed by the pink polygon. The solid black line in the middle is the median curve, which has the largest depth value. All the grey lines in the background are the dataset curves.}
	\end{center}
\end{figure}

In this section, we study how the $50\%$ central region is affected by different choices of depth notions in the given outlier models. 
A good $50\%$ central region is supposed to be compact and not affected by outliers. We compare our depth notion with the modified band depth (MBD)~\citep{Lopez-Pintado2009} and the extremal depth (ED)~\citep{Narisetty2015}. 
When using the total variation depth, the $50\%$ central region is constructed by $50\%$ of the deepest curves after removing the detected shape outliers, as described in \S\ref{subsec:outlierDetectRule}. The resulting central regions clearly differ by choosing different depth notions in Models 3 and 4.
From Fig.~\ref{fig:centralRegion}, we see that the central region constructed using MBD is contaminated by the outliers in Models 3 and 4, and even contains sudden peaks, while ED leads to compact central regions because of its emphasis on extremal properties. The reason MBD fails for Models 3 and 4 is because it only accounts for averaged magnitude outlyingness, and for outliers that only deviate in a short time period as in Models 3 and 4, MBD may assign large depth values, so that these outliers fall into the first half of the deepest curves. By definition, TVD also considers the averaged magnitude outlyingness as MBD. However, by removing the detected shape outliers first, the central region remains compact. All other models show depth notions leading to compact central regions, and we only pick the results for Model 7 in the bottom of Fig.~\ref{fig:centralRegion}. Overall, the performance of our proposed method is appealing in constructing the $50\%$ central region.

\subsection{Outlier detection}\label{subsec:detect}
In this section, we use our outlier detection rule described in \S\ref{subsec:outlierDetectRule} to detect outliers.
\cite{Sun2011} examined the outlier detection performance of the functional boxplot using the modified band depth and $1.5$ times of the $50\%$ central region rule. Later on, \cite{Narisetty2015} compared the outlier detection performance by replacing MBD with their proposed ED in the functional boxplot. \cite{Arribas-Gil2014} proposed the outliergram for shape outlier detection, and magnitude outliers were still detected by the functional boxplot using the modified band depth. All of these methods use the functional boxplot, but with different choices of depth, or they exploit other tools for detecting shape outliers. We now compare the outlier detection performance of the functional boxplot using MBD and ED, the functional boxplot together with the outliergram (OG+MBD), and our proposed method (TVD+MSV), where TVD is used in the functional boxplot for magnitude outliers and shape outlier are detected by MSV.

In the experiment, we assess the performance by the true positive rate (TPR), which is the ratio of the number of correctly detected outliers by the number of true outliers, and the false positive rate (FPR), which is the ratio of the number of wrongly detected outliers by the number of true non-outliers. A larger TPR means more outliers are correctly detected, and a smaller FPR means fewer non-outliers are falsely detected as outliers. The results are shown in Table~\ref{tab:detect}. We see that our proposed outlier detection method is very satisfactory for all models, especially for Models 3 - 7. Even for Model 4, which ED favors, our detection method successfully detects all of the outliers with a TPR of $100\%$, and it outperforms the functional boxplot with ED, for which the TPR is $86.68\%$. For Model 6, where shape outliers only have small oscillations, all the other methods fail with very low TPRs, but our method still gains high accuracy. For shape outliers that also show outlyingness in magnitude as in Models 3, 5 and 7, the outliergram correctly detects most of the outliers, while our method has similar or higher TPRs, but much lower FPRs. For magnitude outliers in Model 2, all the methods perform well. For Model 1, where no outliers exist, all the methods except the outliergram retain a low FPR. In fact, a relatively high FPR of the outliergram is observed for all the models, indicating that the outliergram tends to falsely detect too many outliers.

\begin{table}[ht!]
	\centering
	\caption{\label{tab:detect}Results of outlier detection using different methods for different models. TPR is true positive rate, and FPR is false positive rate. MBD means functional boxplot by MBD. ED means functional boxplot by ED. Outliergram+MBD means detecting shape outliers by outliergram as well as detecting magnitude outliers by functional boxplot. TVD+MSV means our proposed outlier detection rule.}
	\vskip 0.5cm
	\def~{\hphantom{0}}
	\begin{tabular}{c|c|cccc}

	&		&	MBD	&	ED	&	OG+MBD	&	TVD+MSV	\\	\hline	\hline
Model 1	&	FPR	&	0.07(0.28)	&	0.03(0.17)	&	5.42(2.43)	&	0.07(0.28)	\\	\hline	
Model 2	&	TPR	&	99.2(2.99)	&	98.6(5.05)	&	99.2(2.99)	&	99.25(2.78)	\\	\hline	
	&	FPR	&	0.03(0.2)	&	0.01(0.13)	&	4.74(2.36)	&	0.04(0.22)	\\	\hline	
Model 3	&	TPR	&	81.62(14.43)	&	86.68(12.78)	&	87.61(11.57)	&	99.98(0.43)	\\	\hline	
	&	FPR	&	0.05(0.24)	&	0.03(0.17)	&	3.28(2.04)	&	0.05(0.23)	\\	\hline	
Model 4	&	TPR	&	48.53(18.87)	&	86.68(12.38)	&	76.99(18.45)	&	100(0)	\\	\hline	
	&	FPR	&	0.05(0.26)	&	0.02(0.14)	&	3.41(2.11)	&	0.05(0.23)	\\	\hline	
Model 5	&	TPR	&	83.22(14.71)	&	81.62(15.16)	&	99.96(0.7)	&	100(0)	\\	\hline	
	&	FPR	&	0.04(0.23)	&	0.02(0.14)	&	2.21(1.8)	&	0.05(0.25)	\\	\hline	
Model 6	&	TPR	&	0.16(1.5)	&	0.01(0.35)	&	7.35(8.62)	&	99.73(4.05)	\\	\hline	
	&	FPR	&	0.06(0.27)	&	0.02(0.17)	&	4.81(2.42)	&	0.04(0.24)	\\	\hline	
Model 7	&	TPR	&	46.02(24.7)	&	36.06(22.8)	&	99.96(0.9)	&	100(0)	\\	\hline	
	&	FPR	&	0.04(0.22)	&	0.02(0.13)	&	2.17(1.78)	&	0.04(0.21)	\\	\hline			
	\end{tabular}
\end{table}

\section{Applications}
In this section, we apply our proposed outlier detection procedure and visualization tools to three applications: the sea surface temperature in Ni\~no zones, the sea surface height in the Red Sea, and the surveillance video of a meeting room. The three applications cover examples of datasets consisting of curves, images, and video frames.

\subsection{Sea surface temperature}
The El Ni\~no southern oscillation (ENSO), irregular cycles of warm and cold temperatures in the eastern tropical Pacific Ocean, has a global impact on climate and weather patterns, including temperature, rainfall, and wind pressure. There are several indices to measure ENSO, one of which is the sea surface temperature in the Ni\~no regions. The warm phase of ENSO is referred as El Ni\~no, and the cold phase is called La Ni\~na. For example, the years 1982--1983, 1997--1998, and 2015--2016, were reported as strong El Ni\~no years, because the sea surface temperature anomalies were significantly larger for several consecutive seasons.

\begin{figure}[h]
	\begin{center}
		\includegraphics[width=\textwidth]{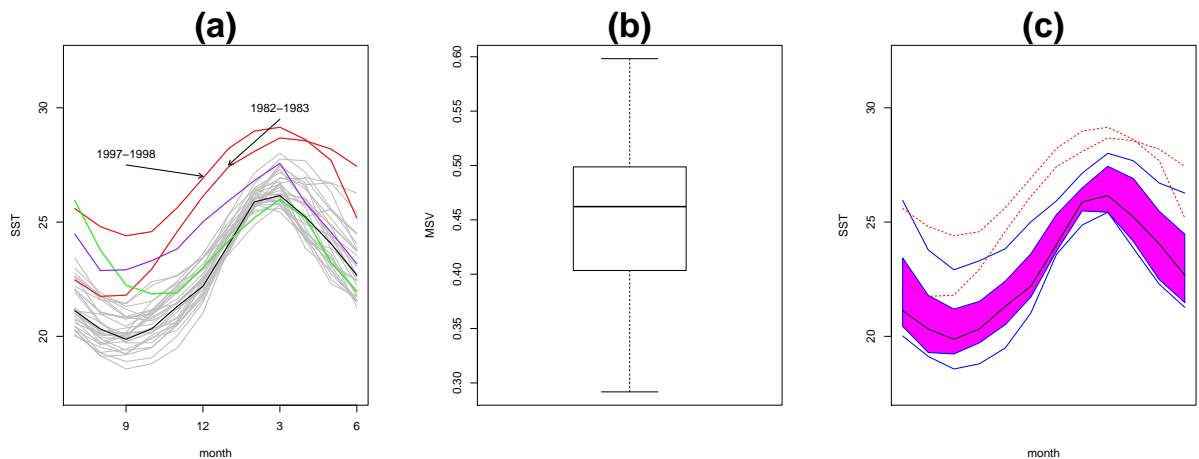}
		\caption{\label{fig:SST}The sea surface temperatures in a Ni\~no zone from July 1982 to June 2016. (a) The original data. (b) The boxplot of the modified shape variation. (c) The functional boxplot with the total variation depth.}
	\end{center}
\end{figure}

In this application, we use our method on the monthly sea surface temperature in one of the Ni\~no zones ($0-10^\circ$ South and $90-80^\circ$ West) from July 1982 to June 2016. In Fig.~\ref{fig:SST}~(a), each curve represents a yearly sea surface temperature from July to next June. There are $35$ curves in total. We see in Fig.~\ref{fig:SST}~(b) that there are no shape outliers, and two magnitude outliers have been detected, which are the years 1982--1983 and 1997--1998. The outliers coincide with the strong El Ni\~no events. The upper envelop shown as the upper blue line in Fig.~\ref{fig:SST}~(c), partly comprises July and August of 1983 (shown by the green line in Fig.~\ref{fig:SST}~(a)), the extension of the El Ni\~no year 1982--1983, and September to December of 2015 (shown by the purple line in Fig.~\ref{fig:SST}~(a)), agreeing with the El Ni\~no year 2015--2016. The median is the year 1989--1990 shown as the black line in Fig.~\ref{fig:SST}~(a), when no obvious El Ni\~no event occurred.

\subsection{Sea surface height}
\begin{figure}[ht!]
	\begin{center}
		\includegraphics[width=\textwidth]{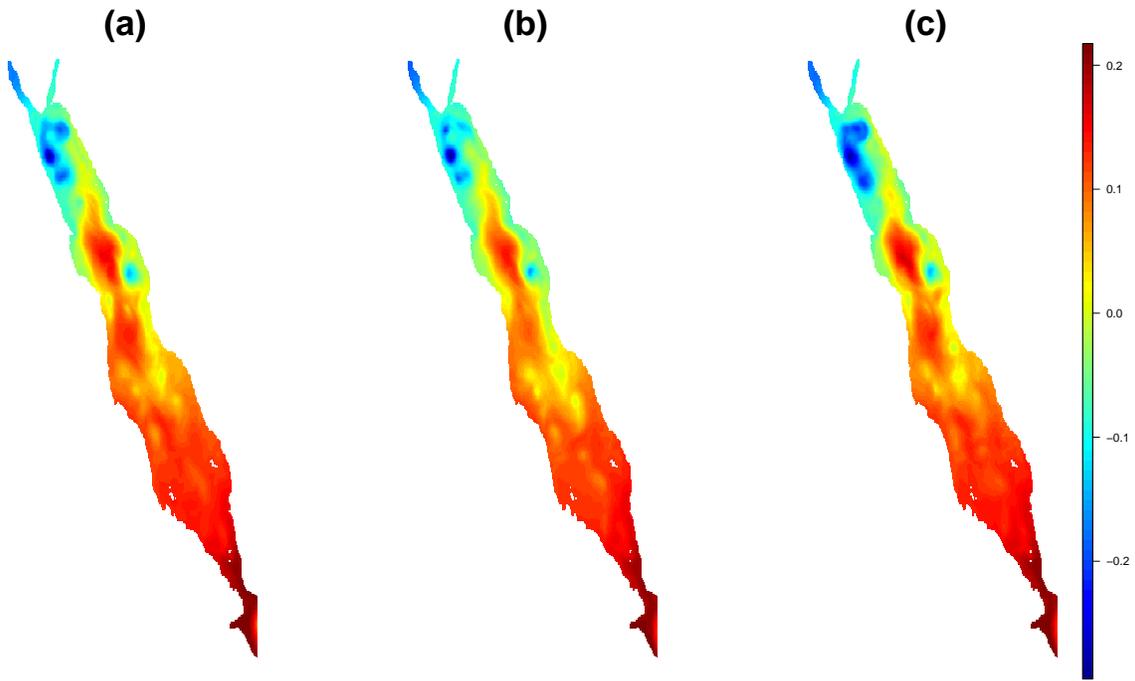}
		\caption{\label{fig:SSH}Runs of the sea surface height model. (a) The median. (b) One of the shape outliers. (c) One of the magnitude outliers.}.
		\end{center}
	\end{figure}

	Sea surface height plays an important role in understanding ocean currents. Geophysical scientists run simulations with different initial conditions to generate ensembles. To understand the variability in these model runs, we apply our method to $50$ simulated sea surface temperatures of the Red Sea on January 1, 2016, where each observation is an image with 36,004 valid values.
	The shape outlyingness plot detects two shape outliers using the factor of $1.5$ in the boxplot. As an example, one shape outlier is shown in Fig.~\ref{fig:SSH}~(b). After removing the two shape outliers, the functional boxplot is applied using the total variation depth. One of the seven magnitude outliers is shown in Fig.~\ref{fig:SSH}~(c), and the median is shown in Fig.~\ref{fig:SSH}~(a).
	We see that compared to the median, the shape outlier has a different pattern in the northern part of the Red Sea, and the magnitude outlier has extreme heights in some places in the middle part of the Red Sea.

\subsection{Surveillance video}
	Video datasets consist of a sequence of video frames, each of which can be treated as one functional observation. \cite{Hubert2016} illustrated their outlier detection method using a beach surveillance video. We explore a video of a meeting room, filmed by \cite{Li2004}, and available at \url{http://perception.i2r.a-star.edu.sg/bk_model/bk_index.html}. 
	In this video, there was nobody in the meeting room at the beginning, and only a curtain was moving with the wind. Later, a person came in, stayed for a while, and left, three times. The goal is to detect those video frames, where either the curtain is moving too far away, or the person is in the meeting room.
	
\begin{figure}[h]
		\begin{center}
			\includegraphics[width=\textwidth]{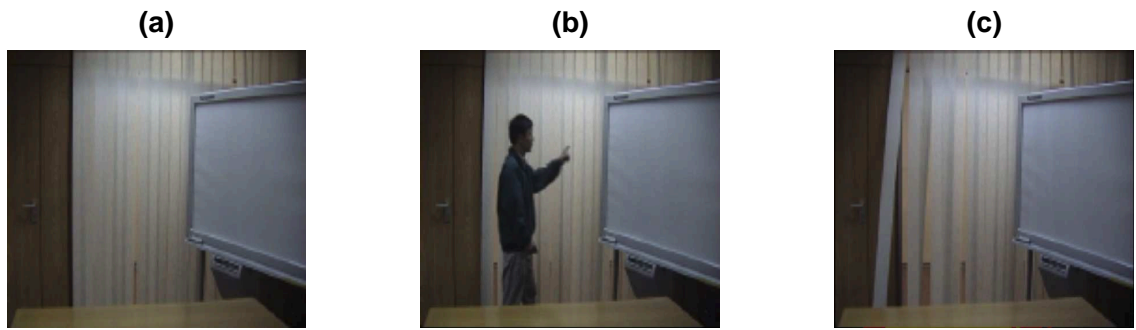}
			\caption{\label{fig:video}The video frames. (a): the median; (b): one of the shape outliers; and (c): one of the magnitude outliers.}
		\end{center}
	\end{figure}

	This video was filmed for 94 seconds, and consists of 2,964 frames. We first equally sub-sampled 1,280 out of 20,480 pixels in each video frame to capture the main features of the image, and then apply our outlier detection method. Our method detects 306 shape outliers and 110 magnitude outliers. The median in Fig.~\ref{fig:video}~(a) shows the most representative frame during the 94 seconds. It indicates the typical position of the curtain in the video with nobody in the meeting room. Fig.~\ref{fig:video}~(b) is one example of the shape outliers, with a person in the meeting room. Fig.~\ref{fig:video}~(c) is one example of the detected magnitude outliers, where we see the curtain moved far away from the median. The detected outlying frames cover the periods in the video where the person was inside the meeting room.
	
\section{Discussion}
In this paper, we proposed a new depth notion, the total variation depth, for functional data. 
We illustrated that this model has many attractive properties and in particular, we highlight its decomposition as especially useful for detecting shape outliers of the most challenging type.
We proposed the modified shape variation defined via the decomposition as an indicator of the shape outlyingness and constructed the shape outlyingness plot to detect shape outliers. Together with the functional boxplot, our proposed outlier detection procedure can detect both magnitude and shape outliers. Good performances of outlier detection were demonstrated through simulation studies with various types of outlier models. Moreover, we also developed a set of visualization tools for functional data, where we display the original observations with informative summary statistics, such as the median and the central region, and highlight the detected magnitude and shape outliers, along with the corresponding shape outlyingness plot and the functional boxplot.

Another advantage of the proposed total variation depth is that it can be easily extended to multivariate functional data by properly defining the indicator function $R_{f}(t)=\mathbbm{1}\{ X(t)\leq f(t)\}$ in Definition~\ref{def:pointTVD}. For example, for a multivariate function $\mathbf{f}(t)$ that takes values in $\mathbb{R}^p$, let  $R_{\mathbf{f}}(t)=\mathbbm{1}\{\mathbf{f}(t)\in \mathbf{S}(t)\}$, where $\mathbf{S}(t)$ is a properly defined set for the multivariate stochastic process $\mathbf{X}$ at time $t$. One naive choice would be $\mathbf{S}(t)=\{\mathbf{X}(t):\|\mathbf{X}(t)\|\leq\|\mathbf{f}(t)\|\}$. However, further research is needed for the properties and outlier detection performances of the multivariate total variation depth.

\newpage
\renewcommand{\theequation}{A.\arabic{equation}}
\setcounter{equation}{0}  
\section*{Appendix}
{\bf Estimation of the total variation depth and shape variation}

Recall that the pointwise total variation depth is defined as 
$D_f(t)=p_f(t)\{1-p_f(t)\},$ where 
$p_f(t)=\mathbb{P}\{ X(t)\leq f(t)\}$.
Given $n$ observations at $m$ time points $X_j(t_i)$, $j=1,\ldots,n$, $i=1,\ldots,m$, we estimate $D_f(t_i)$ by 
$\hat D_f(t_i)=\hat p_f(t_i)\{1-\hat p_f(t_i)\},$ where 
$\hat p_f(t_i)=\#\{j:X_j(t_i)\leq f(t_i)\}/n$, the proportion of $X_j(t_i)$'s that are below $f(t_i)$. To estimate the shape variation, which is defined by  $S_f\{t;s(t)\}=
\hbox{Var}\big(\mathbb{E}[R_f(t)\mid R_f\{s(t)\}]\big)/D_f(t)$ for $D_f(t)\neq0$, namely, $p_f(t_i)\neq 1$, we show how to estimate $\hbox{Var}\big(\mathbb{E}[R_f(t)\mid R_f\{s(t)\}]\big)$ for two consecutive time $t_i$ and $t_{i-1}$ when $i>1$. Since
\[\mathbb{E}\{R_f(t)\mid R_f(s)\}=\mathbb{P}\{R_f(t)=1\mid R_f(s)\}=
\mathbb{P}\{X(t)\leq f(t)\mid R_f(s)\},\] then
\[
\begin{array}{rcl}
\hbox{Var}[\mathbb{E}\{R_f(t_i)\mid R_f(t_{i-1})\}]&=&\hbox{Var}[\mathbb{P}\{X(t)\leq f(t_i)\mid R_f(t_{i-1})\}]\\
&=&\mathbb{E}[\mathbb{P}^2\{X(t_i)\leq f(t_i)\mid R_f(t_{i-1})\}]-\mathbb{E}^2[\mathbb{P}\{X(t_i)\leq f(t_i)\mid R_f(t_{i-1})\}].\\
\end{array}
\]
We estimate the second term
$
\mathbb{E}^2[\mathbb{P}\{X(t_i)\leq f(t_i)\mid R_f(t_{i-1})\}]
$ by $\hat p^2_f(t_i)$, and we show the first term can be simplified as follows:
\[
\begin{array}{rcl}
\mathbb{E}[\mathbb{P}^2\{X(t_i)\leq f(t_i)\mid R_f(t_{i-1})\}]&=&\mathbb{P}\{R_f(t_{i-1})=0\}\mathbb{P}^2\{X(t_i)\leq f(t_i)\mid R_f(t_{i-1})=0\}\\
&&+\mathbb{P}\{R_f(t_{i-1})=1\}\mathbb{P}^2\{X(t_i)\leq f(t_i)\mid R_f(t_{i-1})=1\}\\
&=&\mathbb{P}\{X(t_{i-1})\leq f(t_{i-1})\}\mathbb{P}^2\{X(t_i)\leq f(t_i)\mid X(t_{i-1})\leq f(t_{i-1})\}\\
&&+\mathbb{P}\{X(t_{i-1})> f(t_{i-1})\}\mathbb{P}^2\{X(t_i)\leq f(t_i)\mid X(t_{i-1})> f(t_{i-1})\}.\\
\end{array}
\]

Let \[\hat p_f(t_i,t^-_{i-1})=\#\{j:X_j(t_i)\leq f(t_i), X_j(t_{i-1})\leq f(t_{i-1})\}/n,\] and \[\hat p_f(t_i,t^+_{i-1})=\#\{j:X_j(t_i)\leq f(t_i), X_j(t_{i-1})> f(t_{i-1})\}/n,\] we then estimate $\mathbb{E}[\mathbb{P}^2\{X(t_i)\leq f(t_i)\mid R_f(t_{i-1})\}]$ by
\[\hat p^2_f(t_i,t^-_{i-1})/\hat p_f(t_{i-1})+\hat p^2_f(t_i,t^+_{i-1})/(1-\hat p_f(t_{i-1})),\] when $p_f(t_{i-1})\neq 1$. 
For $\hat p(t_{i-1})=1$, the second part vanishes. Then, the estimator for $S_f(t_i;t_{i-1})$ is 
\[
\hat S_f(t_i;t_{i-1})=\left\{
\begin{array}{lcl}
1&,&\hat p(t_i)=1,\\
\dfrac{\hat p_f^2(t_i,t^-_{i-1})/\hat p_f(t_{i-1})-\hat p_f^2(t_i)}{\hat p_f(t_i)\{1-\hat p_f(t_i)\}}&,&\hat p(t_i)\neq1,\hat p(t_{i-1})=1,\\
\dfrac{\left\{\dfrac{\hat p_f^2(t_i,t^-_{i-1})}{\hat p_f(t_{i-1})}+\dfrac{\hat p_f^2(t_i,t^+_{i-1})}{1-\hat p_f(t_{i-1})}\right\}-\hat p^2(i_\leq)}{\hat p_f(t_i)\{1-\hat p_f(t_i)\}}&,&\hat p(t_i)\neq1,\hat p(t_{i-1})\neq1.\\
\end{array}\right.
\]

For the weight function $w(t)$, which is the normalized standard deviation of the random function at time $t$, we use the sample standard deviation for estimation. At time $t_i$, the empirical estimator of weight function $w(t_i)$ is given by
\[\hat w(t_i)=\frac{\hat{\hbox{sd}}(t_i)}{\sum_{r=1}^{m}\hat{\hbox{sd}}(t_r)},
\]
where $\hat{\hbox{sd}}(t_i)=\left[\sum^n_{j=1}\left\{X_j(t_i)-\frac{\sum^n_{k=1}X_k(t_i)}{n}\right\}^2/(n-1)\right]^{1/2}.$
Similarly for the weight function $v(t)$ in the shape variation, which is the normalized difference in function values, the empirical estimator at time $t_i$ is 
\[\hat v(t_i)=\frac{\mid f(t_i)-f(t_{i-1})\mid}{\sum_{r=2}^{m}\mid f(t_r)-f(t_{r-1})\mid}.\]
Finally, the estimators of the TVD and SV are given by
\[
\begin{array}{rcl}
\widehat{\hbox{TVD}}(f)&=&\sum_{i=1}^{m}\hat w(t_i)\hat{D}_{f}(t_i),\\
\widehat{\hbox{SV}}(f)&=&\sum_{i=2}^{m}\hat v(t_i)\hat{S}_{f}(t_i;t_{i-1}).\\
\end{array}
\]

{\bf Proof of Theorem~\ref{theorem:consistency}}

First, we provide the detail of the interpolating continuous process $\tilde X$, which is a linear interpolation of the discrete observations in C5.
Suppose we have $n$ samples $\{X_1(t), \cdots, X_n(t)\}$ observed at $m$ points $\{t_1<t_2<\ldots<t_m\}.$  The sample $\tilde X_n(t)$ is given by
\[
\tilde X_n(t)=\left\{
\begin{array}{lcl}
X_n(t_i)\dfrac{t_i+t_{i+1}-2t}{t_{i+1}-t_i}+\bar X(t_i)\dfrac{2(t-t_i)}{t_{i+1}-t_i}&,&t\in[t_i,\dfrac{t_i+t_{i+1}}{2}]\\
X_n(t_{i+1})\dfrac{2t-t_i-t_{i+1}}{t_{i+1}-t_i}+\bar X(t_i)\dfrac{2(t_{i+1}-t)}{t_{i+1}-t_i}&,&t\in[\dfrac{t_i+t_{i+1}}{2},t_{i+1}]\\
\end{array}\right.,
\]
where $\bar X(t_i)=\sum_{j=1}^{n}X_j(t_i)$.

Now, consider our proposed $\hbox{TVD}(f)$ and its estimator $\widehat{\hbox{TVD}}(f)$. To prove that $\widehat{\hbox{TVD}}(f)$ is consistent under conditions C1 - C5, \cite{Claeskens2014} have shown that we only need to show $\sup_{f(t)\in\mathbb{R}}|D_f(t)-\hat D_f(t)|\rightarrow0$, a.s. as $n\rightarrow\infty$.
Note that $D_f(t_i)=p_f(t_i)\{1-p_f(t_i)\}$ is the simplicial depth for the univariate case up to a constant, and \cite{Liu1990} has proved that $\sup_{f(t)\in\mathbb{R}}|D_f(t)-\hat D_f(t)|\rightarrow0$, a.s. as $n\rightarrow\infty$ under condition C6. This ends the proof.

\newpage
\bibliographystyle{chicago}
\bibliography{../ref/reference}

\end{document}